# *p*-norm-like Constraint Leaky LMS Algorithm for Sparse System Identification

FENG Yong[1], ZENG Rui[2], WU Jiasong[2]

1. School of Biological Science and Medical Engineering, 2. School of Computer Science and Engineering
Southeast University
Nanjing, Jiangsu, 210096, China
(email: fengyong, zengrui, jswu@seu.edu.cn)

**Abstract:** In this paper, we propose a novel leaky least mean square (leaky LMS, LLMS) algorithm which employs a *p*-norm-like constraint to force the solution to be sparse in the application of system identification. As an extension of the LMS algorithm which is the most widely-used adaptive filtering technique, the LLMS algorithm has been proposed for decades, due to the deteriorated performance of the standard LMS algorithm with highly correlated input. However, both of them do not consider the sparsity information to have better behaviors. As a sparse-aware modification of the LLMS, our proposed $\ell_p$like-LLMS algorithm, incorporates a *p*-norm-like penalty into the cost function of the LLMS to obtain a shrinkage in the weight update, which then enhances the performance in sparse system identification settings. The simulation results show that the proposed algorithm improves the performance of the filter in sparse system settings in the presence of noisy input signals.

**Key words:** Adaptive filtering; *p*-norm-like constraint; sparse system identification; leaky least mean square algorithm; LLMS; LMS

## 1. Introduction

A sparse system is a system whose impulse response contains very few nonzero coefficients, while other taps are zeros or nearly zeros. Practically, Sparse systems are very common, especially in communications, for instance, digital TV transmission channels, sparse sparse acoustic echo paths, sparse wireless multi-path channels, and so forth. The least mean square (LMS) algorithm is one of the most widely-used techniques among the traditional adaptive filtering algorithms for system identification, owning to its computational simplicity, efficiency and robustness [1]. For the application of system identification, however, the standard LMS does not assume any structural information about the unknown system, which in turn, results a deteriorated performance in terms of the steady-state excess mean square error (EMSE) and the convergence speed [2]. Moreover, the performance of the standard LMS also deteriorates when the input of the system is highly correlated [3]. Thus, numerous modifications of the standard LMS algorithm have been proposed to deal with this problem, and among the famous LMS variants is the leaky LMS (LLMS) algorithm [3].

As one of the two important families of sparse LMS algorithms, some new LMS algorithms with different norm constraints have been proposed, for example, the $\ell_1$-norm constraint LMS ($\ell_1$-LMS) [2, 4], $\ell_0$-norm constraint LMS ($\ell_0$-LMS) [5, 6], $\ell_p$-norm constraint LMS ($\ell_p$-LMS) [7, 8] and $\ell_p$like-norm constraint LMS ($\ell_p$like-LMS) [9,10] in which the corresponding $\ell_1$, $\ell_0$, $\ell_p$ and $\ell_p$-like norms are incorporated into the cost function of the standard LMS algorithm, respectively, aiming at increasing the convergence rate as well as decreasing the mean square error (MSE) at the same time.

To combine the advantages of both the sparse LMS and LLMS algorithm, we propose a *p*-norm-like constraint LLMS algorithm ($\ell_p$like-LLMS), in which a *p*-norm-like constraint is embedded into the cost function of the LLMS to force the solution to be sparse in the application of system identification.

This paper is organized as follows. In Section 2, the standard LMS and LLMS algorithms are briefly reviewed and the proposed algorithm $\ell_p$like-LLMS is then derived. In Section 3, simulation results are shown which compare the performance of the proposed algorithm with those of the standard LMS, standard LLMS and $\ell_p$like-LMS algorithm in sparse system identification settings. Finally, this paper is concluded in Section 4.

## 2. Algorithms

Throughout this paper, we apply the following notations. Matrices and vectors are denoted by boldface upper-case letters and boldface lower-case letters, respectively. And variables and constants are in italic lower-case letters. The superscripts $(\cdot)^T$ represents the transpose operators, and $E[\cdot]$ is the expectation operator.

Let $y_k$ be the output of an unknown system with an additional noise $n_k$ at time $k$, which can be written as the following by a linear model

$$y_k = \mathbf{w}^T \mathbf{x}_k + n_k, \qquad (1)$$

where the weight vector $\mathbf{w}$ of length $N$ denotes the impulse response (IR) of the unknown system, $\mathbf{x}_k = [x_k, x_{k-1}, \cdots, x_{k-N+1}]^T$ represents the input signal with covariance $\mathbf{R}$ and $n_k$ is a stationary noise with zero mean and variance $\sigma_k^2$. Given the input $\mathbf{x}_k$ and output $y_k$ following the settings above, the standard LMS algorithm is applied to estimate the vector $\mathbf{w}$ with the cost function $J_k$ defined as

$$J_k = e_k^2 / 2, \qquad (2)$$

where $e_k = y_k - \mathbf{w}_k^T \mathbf{x}_k$ denotes the instantaneous error and $\mathbf{w}_k = [w_{k,1}, w_{k,2}, \cdots, w_{k,N}]^T$ represents the estimated weight vector of the system at time $k$, note that here "1/2" is employed for the convenience of computation. Thus, the update equation is then written as

$$\mathbf{w}_{k+1} = \mathbf{w}_k - \mu \frac{\partial J_k}{\partial \mathbf{w}_k} = \mathbf{w}_k + \mu e_k \mathbf{x}_k, \qquad (3)$$

where $\mu$ is the step size, satisfying $0 < \mu < \lambda_{max}^{-1}$, where $\lambda_{max}$ is the maximum eigenvalue of the aforementioned matrix $\mathbf{R}$ [1].

In order to conquer the weight drift problem in the standard LMS algorithm under the condition that the input signal is highly correlated, the LLMS algorithm [2] is introduced, which can improve the convergence and stability at the same time. The LLMS algorithm takes a leakage factor $\gamma(0 < \gamma < 1)$ to control the weight update of the LMS algorithm, and its cost function is defined as

$$J_k = e_k^2 / 2 + \gamma \|\mathbf{w}_k\|_2^2. \qquad (4)$$

And the weight vector **w** is then updated by

$$\mathbf{w}_{k+1} = (1 - \mu\gamma)\mathbf{w}_k + \mu e_k \mathbf{x}_k. \qquad (5)$$

For sparse system identification, the $\ell_p$like-LMS [9] has been proposed with the new cost function $J_{k,pl}$ represented by

$$J_{k,pl} = e_k^2 / 2 + \gamma_{pl} \|\mathbf{w}_k\|_{pl}, \qquad (6)$$

where the $p$-norm-like is defined as $\|\mathbf{w}_k\|_{pl} \triangleq \sum_{i=1}^{N} |w_{k,i}|^p$ with $0 < p < 1$, and $\gamma_{pl}$ is a constant controlling the trade-off between the convergence rate and estimation error. Thus, the update of the $\ell_p$like-LMS is then derived as

$$\mathbf{w}_{k+1} = \mathbf{w}_k + \mu e_k \mathbf{x}_k - \rho_{pl} \frac{p \, \text{sgn}(\mathbf{w}_k)}{\varepsilon_{pl} + |\mathbf{w}_k|^{1-p}}, \qquad (7)$$

where $\rho_{pl} = \mu\gamma_{pl}$ weighting the $p$-norm-like constraint, $\varepsilon_{pl}$ is a constant which is used to avoid the denominator of the last term in (7) being zero, and sgn($x$) is the sign function, which is 0 for $x = 0$, 1 for $x > 0$ and -1 for $x < 0$. Note that the sign can also apply to vectors as in this paper, i.e., it "signs" the vector element-by-element.

In this paper, we grasp the idea of applying the $p$-norm-like constraint in the LLMS algorithm, to obtain a new approach named the $\ell_p$like-LLMS, whose cost function is defined as

$$J_k = e_k^2 / 2 + \gamma \|\mathbf{w}_k\|_2^2 + \gamma_{pl} \|\mathbf{w}_k\|_{pl}. \qquad (8)$$

Using gradient descent method, the update for the $\ell_p$-LLMS algorithm is given by

$$\mathbf{w}_{k+1} = (1 - \mu\gamma)\mathbf{w}_k + \mu e_k \mathbf{x}_k - \rho_{pl} \frac{\|\mathbf{w}_k\|_p^{1-p} \text{sgn}(\mathbf{w}_k)}{\varepsilon_{pl} + |\mathbf{w}_k|^{1-p}}. \qquad (9)$$

Note that we employ $(1 + \mu\gamma)$ here instead of $(1 - \mu\gamma)$ to obtain a lower stable error, i.e.,

$$\mathbf{w}_{k+1} = (1 + \mu\gamma)\mathbf{w}_k + \mu e_k \mathbf{x}_k - \rho_{pl} \frac{\|\mathbf{w}_k\|_p^{1-p} \text{sgn}(\mathbf{w}_k)}{\varepsilon_{pl} + |\mathbf{w}_k|^{1-p}}. \qquad (10)$$

Thus, it is not directly extended from the LLMS algorithm.

## 3. Simulations

In this section, measured by the common index the mean square deviation (MSD, defined as $\text{MSD}_k = E[\|\mathbf{w} - \mathbf{w}_k\|_2^2]$), the performance of the $\ell_p$like-LLMS is compared with those of the standard LMS, LLMS and $\ell_p$like-LMS algorithms in following numerical simulations, aiming at showing their tracking and steady-state performance in the sparse system identification settings.

We estimate a sparse unknown system of 16 taps with 1, 4, 8 or 16 taps that are assumed to be nonzeros, which makes the sparsity ratio (SR) be 1/16, 4/16, 8/16 or 16/16, respectively. The positions of nonzero taps are chosen randomly and the values are 1's or -1's randomly. We have 8000 iterations for each sparsity level. The input is a correlated signal generated by $x_{k+1} = 0.8x_k + u_k$ and then normalized to variance 1, where $u_k$ is a white Gaussian noise with variance $10^{-3}$. The observed noise is assumed to be white Gaussian processes of length 8016 with zero mean and variance $10^{-2}$, i.e., the signal noise ratio (SNR) is set to be 20 dB. Other parameters are carefully selected as listed in Table 1. All the simulations are averaged over 200 independent runs to smooth out the MSD curves.

**Table 1**. Parameters of the algorithms in the experiment.

| Algorithms | $\mu$ | $\rho$ | $\varepsilon$ | $p$ | $\gamma$ |
|---|---|---|---|---|---|
| LMS | 0.015 | / | / | / | / |
| $\ell_p$like-LMS |  | 0.003[a] |  |  | / |
| LLMS |  | 0.002[b] | 10 | 0.5 | 0.005[e] |
| $\ell_p$like-LLMS |  | 0.0015[c] |  |  | 0.0005[f] |
|  |  | 0.0001[d] |  |  |  |

[a]~[d] are for SR=1/16, 4/16, 8/16, 16/16, respectively.
[e] is for SR=1/16, 4/16 and 8/16; [f] is for SR=16/16.

Fig. 1 shows the MSD curves of the algorithms tested for different sparsity levels. From Fig. 1, one can see that, for all different sparsity levels except the totally non-sparse case, i.e., SR = 1/16, 4/16 and 8/16, the proposed $\ell_p$like-LLMS always achieves the best performance in term of steady-state MSD. Moreover, when a system is very sparse, i.e., SR = 1/16, both the $\ell_p$like-LMS and $\ell_p$like-LLMS yield lower stable MSDs than those of the LMS and LLMS, due to the function of the $p$-norm-like sparse constraint which attracts the taps of the impulse response to zeros. However, as the sparsity ratio increases, the $\ell_p$like-LMS may performs worse than the standard LMS algorithm, which is expected, since the $\ell_p$like-LMS is developed to deal with sparse cases. But our $\ell_p$like-LLMS still performs the best, owning to its feature which combines the advantages of both the standard LLMS and the $\ell_p$like-LMS. In the totally-non-sparse case, that is, SR = 16/16, we may decease the constraint weight parameter $\rho$ to get the performances of the $\ell_p$like-LMS and $\ell_p$like-LMS almost equal to those of the LMS and LLMS. To sum up, our proposed $\ell_p$like-LLMS outperforms the standard LMS, LMS and $\ell_p$like-LMS for different sparsity levels of the system in sparse system identification settings.

## 4. Conclusions

We propose the $\ell_p$like-LLMS algorithm which incorporates a $p$-norm-like constraint into the cost function of the leaky LMS to force the solution to be sparse in the application of system identification, thus it combines the advantages of both the leaky LMS and the $\ell_p$like-LMS to achieve better performance while the input of the system is highly correlated. And simulations show that our proposed $\ell_p$like-LLMS outperforms the standard LMS, LMS and $\ell_p$like-LMS for different sparsity levels of the system in sparse system identification settings in the presence of noisy input signals.

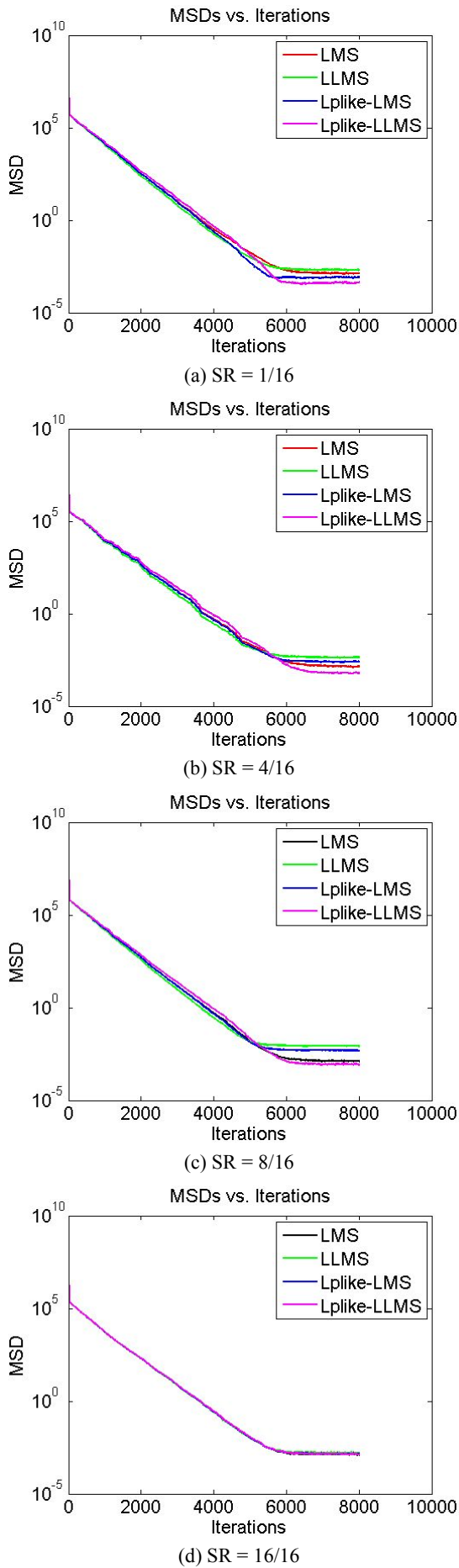

**Fig. 1**. MSD curves of different algorithms with different SRs

## Acknowledgment

This work was supported by the National Basic Research Program of China under Grant 2011CB707904, by the NSFC under Grants 61201344, 61271312, 11301074, and by the SRFDP under Grants 20110092110023 and 20120092120036, the Project-sponsored by SRF for ROCS, SEM, and by Natural Science Foundation of Jiangsu Province under Grant BK2012329 and by Qing Lan Project.